# Tidal Fields and Structure Formation

Rien van de Weygaert

*Kapteyn Astronomical Institute, University of Groningen, P.O. Box 800, 9700 AV, Groningen, the Netherlands*

**Abstract.** The role of tidal shear in the formation of structure in the Universe is explored. To illustrate the possible and sometimes dramatic impact of tidal fields we focus on the evolution of voids. We firstly analyze the role of tidal fields both in the highly symmetric situation of an initially homogeneous ellipsoidal underdensity embedded in an artificially imposed tidal field. In addition, we present selfconsistent case studies consisting of N-body simulations that start from constrained Gaussian initial conditions in which the matter distribution is appropriately sculpted to yield an a priori specified tidal field. We conclude that voids may indeed be induced to collapse. Also, we present evidence for the strong relation between tidal fields and filaments in the mass distribution.

## 1. Introduction

Observations of the galaxy distribution in the universe show that galaxies trace out a frothy pattern, delineated by thin high-density ridges, plateaus and aggregates of galaxies interwoven by a network of practically empty voids. It is generally believed that the structures delineated by the galaxies have formed by gravitational amplification of initially small Gaussian random density fluctuations. One of the issues within this framework is the origin of the highly anisotropic pattern of the characteristically flattened and elongated overdense structures. The anisotropy of these mass concentrations suggests that tidal forces, induced by the inhomogeneous mass distribution itself, are likely to be instrumental in the formation and shaping of the foamlike pattern. Although tidal fields are capable of profoundly influencing structure formation in the universe their role has not yet been explored to any great extent. Here we are specifically interested in the issue of how tidal fields induced by the external inhomogeneities can affect the dynamics and evolutionary fate of voids (van de Weygaert & Babul 1995).

## 2. Homogeneous Ellipsoidal Voids

To address the question of whether voids, in 'Eulerian' space, can collapse due to tidal fields we first explore a highly symmetric situation, the evolution of an underdense homogeneous ellipsoid under the influence of an artificially imposed external tidal field. In this approximation the void is considered to be a homogeneous triaxial ellipsoidal configuration of uniform density $\rho_e$, embedded in a homogeneous background with density $\bar{\rho}$ (see e.g. Icke 1984). The evolution of an isolated homogeneous ellipsoid proceeds through a series of uniform ellipsoids. The only factors that enter into the problem are the initial axis ratios, the initial density, the corresponding initial velocity field, and its cosmological background.

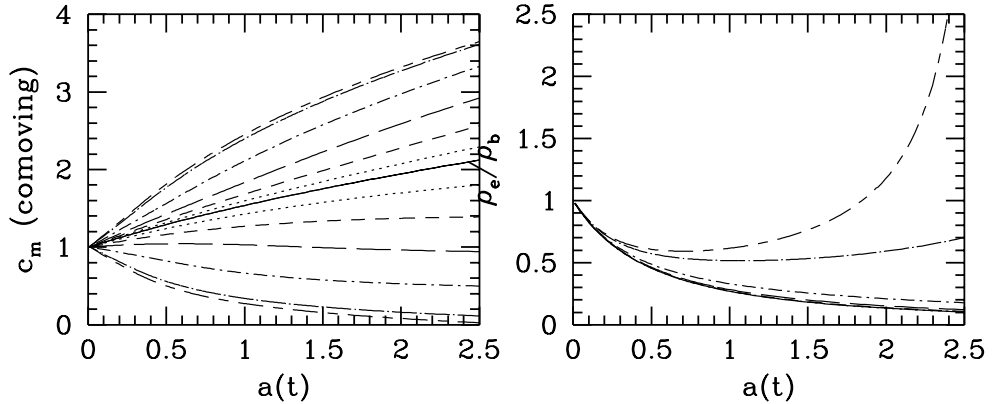

Figure 1. The evolution of (comoving) axis (left) and density (right) of an initially underdense ellipsoid. See text for details.

Within a general field of density perturbations, the ellipsoidal approximation applies mainly in the immediate vicinity of density maxima and minima, where the leading terms in the gravitational potential are the quadratic ones. Numerical simulations (e.g. van de Weygaert & van Kampen 1993) show that in the interior of voids the ellipsoidal approximation may be a good one, as the density profile in the central region of a void evolves into one of nearly uniform density while it expands and gets drained. Moreover, nearly all small-scale structure initially within the protovoid gets destroyed. On the other hand, the applicability of the homogeneous ellipsoid approximation may be limited by the effect of the void on the smooth background, which will get disturbed by the potential disturbance represented by the ellipsoid and by the fact that the void will sweep up matter into (anisotropic) ridges. The resulting background perturbations may engender tidal forces that act back onto the void itself. We assume that at least for the inner region of the homogeneous ellipsoids the above effects can be neglected. In our investigation we impose an artificial tidal field by means of an extra quadratic term in the gravitational potential. Thereby we implicitly assume that the external structures that give rise to the tidal field are located at large distances from the void, and that the field does not vary significantly over the expanse of the void. We also limit ourselves to a spherical initial configuration, so that the principal axes of the external tidal tensor will always be oriented along the principal axes of the mass tensor. The problem then reduces to that of three functions $R_m(t)$, the scale factors of the principal axes of the ellipsoid, $c_m(t) = R_m(t)c_{m,i}$, whose equations of motion are given by

$$\frac{d^2 R_m}{dt^2} = -2\pi G \left[\alpha_m \rho_e + (\frac{2}{3} - \alpha_m)\bar{\rho}\right] R_m - E_{mm} R_m, \qquad (1)$$

where $E_{mm}$ is one of the diagonal traceless tidal tensor components (see equation 3 for the formal definition of the traceless tidal tensor components $E_{mn}$). In Figure 1 we illustrate the evolution of the (comoving) axes $c_m(t)$ (lefthand frame) and density $\rho_e/\rho_b$ of an initially spherical void with a density deficit $\delta_i = -0.02$ at the initial expansion factor $a_i = 0.01$, embedded in an Einstein-de Sitter background universe with Hubble paramter $H_0 = 50$ km/s/Mpc. The velocity perturbation represented by the initial void is taken to correspond to the growing mode solution of linear perturbation theory. The solid line in both frames represents the evolution of the void in the absence of any external influence, functioning as reference configuration. Evidently, the void remains spherical and expands equally fast along all three axes, while the density gradually decreases towards $\rho_e = 0.0$. The non-solid lines correspond to the configurations wherein the void was embedded in an external tidal field.

This field was taken to axisymmetric with initial components $E_{mm}/\frac{3}{4}\Omega H^2 = (-E, -E, 2E)$ and assumed to evolve according to linear theory up to a maximum value. From the dotted lines to the short-long dashed lines the value of $E$ increases from 0.1 to 2.0. Evidently, the presence of the external tide causes a qualitatively different evolution. Relative to the spherical void, the voids expand more rapidly along the $x$ and $y$ axes due to the extra accelerating effect of the negative tidal components $E_{11}$ and $E_{22}$. This extra expansion increases as the external tidal field is stronger. In contrast, the positive tidal force along the $z$ axis causes the void to slow down the expansion along this axis. For the relatively weak external tidal fields this consists of a mere slow-down of expansion along the $z$-axis. The stronger external tidal fields lead to a greater rate of slowing down this expansion. However, even in the intermediate cases the (physical) axis keeps on expanding, and the density of the void keeps on decreasing, so that the effect of the tidal field seems to be restricted to elongating the region. However, in the most extreme cases the external fields are so strong that the $z$ axis starts to collapse. In these cases we see that the density, after an initial decrease, starts to grow again at a certain stage, finally evolving towards $\infty$. In other words, the void has collapsed under the influence of the external tidal field. We should note however that the evolution sketched above is critically dependent on the exact time evolution of the external tidal field.

## 3. Constrained N-body simulations

For a fully self-consistent study of void and tidal field evolution we address the evolution of a void forming in a random density field developing from Gaussian initial conditions. Rather than studying one or more general N-body simulations our strategy consists of investigating a few case studies, based on systematically specified initial conditions. Implicitly we assume that in a field of random density fluctuations voids evolve from dips in the initial conditions (also see van de Weygaert & van Kampen 1993). For the purpose of the simulations, we assume an $\Omega = 1$ universe, a Hubble constant of $H_0 = 50$ km/s/Mpc, and adopt the standard cold dark matter spectrum to characterize the initial density fluctuations — normalized such that $\sigma_0(8h^{-1} \text{Mpc}) = 1.0$ at the present epoch ($a = 1$).

Basically, we explore the evolution of the matter distribution associated with and in the neighbourhood of a dip in the initial density field. Apart from being able to determine the location $\mathbf{r}_{dp}$, the Gaussian scale and the depth of the dip, we sculpt the total matter distribution such that this object is subject to a desired amount of net gravitational and tidal forces. In order to generate these constrained Gaussian density fields, we use the prescription of Hoffman & Ribak (1991), as implemented by van de Weygaert & Bertschinger (1995) for the specific case of peaks. In addition to its scale and location, a dip (or peak) in the smooth density field is characterized by 10 constraints that specify the density field in the immediate vicinity of the dip, and by 8 constraints to specify the 3 components of the smoothed peculiar gravitational acceleration $\mathbf{g}_G(\mathbf{r}_{dp})$ and the 5 independent components of the traceless tidal shear tensor,

$$\mathbf{g}_G(\mathbf{r}_{dp}) = -\nabla_{\mathbf{r}}\phi_G = \frac{3\Omega_0 H_0^2}{8\pi} \int d\mathbf{y} \int d\mathbf{x}\, \delta(\mathbf{x}) W_G(\mathbf{x},\mathbf{y}) \frac{(\mathbf{y}-\mathbf{r}_{dp})}{|\mathbf{y}-\mathbf{r}_{dp}|^3}, \quad (2)$$

and

$$E_{G,mn}(\mathbf{r}_{dp}) = \left. \frac{\partial^2 \phi_G}{\partial r_m \partial r_n} - \frac{1}{3}(\nabla^2 \phi_G)\delta_{mn} \right|_{\mathbf{r}=\mathbf{r}_{dp}}, \quad (3)$$

where $\phi_G$ is the perturbed part of the potential and $\delta(\mathbf{x})$ is the corresponding (unsmoothed) density fluctuation field, while $W_G(\mathbf{x},\mathbf{y})$ is the Gaussian smoothing filter. The above constraints are linear functionals on the density field $\delta(\mathbf{r})$. The constraints constitute a set of $M$ linear functionals $C_m[\delta]$ on $\delta$ with the value $c_m$ at the location of the peak. These linear constraints can be cast in the form of a convolution between $\delta(\mathbf{x})$ and some kernel $H_m(\mathbf{x};\mathbf{r}_{dp})$,

$$C_m[\delta;\mathbf{r}_{dp}] = \int d\mathbf{x}\, \delta(\mathbf{x}) H_m(\mathbf{x},\mathbf{r}_{dp}) = \int \frac{d\mathbf{k}}{(2\pi)^3} \widehat{\delta}(\mathbf{k}) \widehat{H}_m^*(\mathbf{k}) = c_m, \quad (4)$$

where $\widehat{\delta}(\mathbf{k})$ and $\widehat{H}_m(\mathbf{k})$ are the Fourier transforms of $\delta(\mathbf{x})$ and $H_m(\mathbf{x};\mathbf{r}_{dp})$. The kernels for the peculiar gravity and the tidal field constraints, for example, are

$$\widehat{H}_{\mathbf{g}}(\mathbf{k}) = -i\mathcal{H}\frac{\mathbf{k}}{k^2}\widehat{W}(\mathbf{k})e^{i\mathbf{k}\cdot\mathbf{r}_{dp}}, \quad \widehat{H}_{Emn}(\mathbf{k}) = \mathcal{H}\left(\frac{k_m k_n}{k^2} - \frac{1}{3}\delta_{mn}\right)\widehat{W}(\mathbf{k})e^{i\mathbf{k}\cdot\mathbf{r}_{dp}}, \quad (5)$$

where the constant $\mathcal{H} = \frac{3}{2}\Omega_0 H_0^2$ and $\widehat{W}(\mathbf{k})$ is the Fourier transform of the Gaussian filter. The constrained fluctuation field $\delta(\mathbf{r})$ can then be constructed from the Fourier sum (van de Weygaert & Bertschinger 1995)

$$\delta(\mathbf{r}) = \int \frac{d\mathbf{k}}{(2\pi)^3} \left[\widehat{\widetilde{\delta}}(\mathbf{k}) + P(k)\widehat{H}_m(\mathbf{k})\xi_{mn}^{-1}(c_n - \widetilde{c}_n)\right] e^{-i\mathbf{k}\cdot\mathbf{r}}, \quad (6)$$

where $\widetilde{\delta}(\mathbf{x})$ is an unconstrained realization of the fluctuation field with the same power spectrum $P(k)$ as that characterizing constrained field $\delta(\mathbf{x})$, $\widetilde{c}_n$ is the value of the linear functional $C_n$ for the field $\widetilde{\delta}(\mathbf{x})$, and the random Gaussian variate $\widehat{\widetilde{\delta}}(\mathbf{k})$ is its Fourier transform. Also, $\xi_{mn}$ is the constraint's correlation function. In the study presented here the generated constrained fields are sampled onto a $64^3$ grid in a periodic box of length $25h^{-1}$Mpc, transformed into positions and growing mode velocities of $64^3$ particles. These are evolved using a P$^3$M N-body code

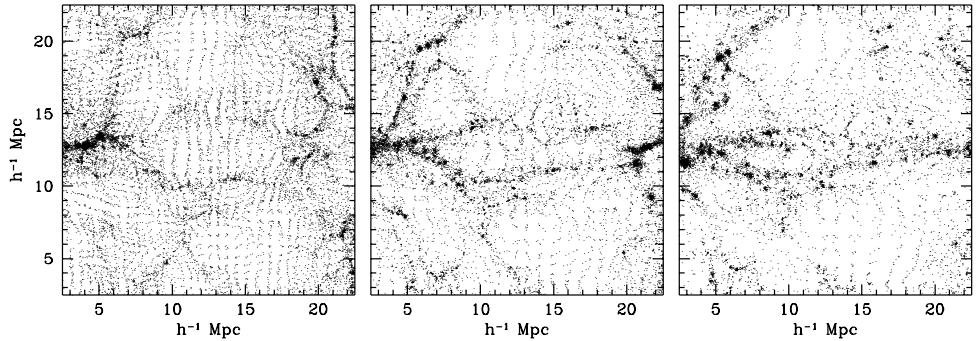

Figure 2. Three timesteps, $a = 0.15$, $0.30$ and $0.45$, in the evolution of a 'collapsing' CDM void. See text for details.

## 4. Results of N-body simulations

In Figure 2 we show a sequence of 3 timesteps ($a = 0.15$, $0.30$ and $0.45$) in the evolution of a void defined on a Gaussian scale of $R_G = 1.25 h^{-1}$ Mpc. It concerns the particle distribution in a $2.0 h^{-1}$ Mpc thick slice through the centre of the simulation box. In the initial density field the dip is at rest and subject to tidal stresses that were quantified via the traceless shear components $\sigma_{mn}$ (in the linear regime, $\sigma_{mn} \propto E_{mn}$). The traceless shear tensor $\sigma_{mn}$ at the location of the density depression is oriented so that the off-diagonal terms are zero, while $\sigma_{xx} = 200$ km/s/Mpc (stretch) and $\sigma_{yy} = \sigma_{zz} = -100$ km/s/Mpc (contraction). These values represent 2–4 $\sigma$ fluctuations in the tidal field. The quadrupolar mass distribution that induces these tidal stressed consists of mass concentrations immediately to the left and right of the void, and large underdensities above and below it. Even though the conditions were set up such that a sphere centered on the centre of the box and with a radius smaller or equal to $7.5 h^{-1}$ Mpc would not be overdense, we do observe that the initial underdensity evolves into a void that is shrinking in the $y$-direction, while it is stretching along the $x$-direction. In other words, the tidal stress leads to the collapse of the void !

As such a collapse of a void under the influence of a tidal field is critically dependent on the evolution and structure of the tidal field itself, we subsequently turn to a description of the tidal field in an arbitrary realization of matter distribution in the CDM scenario. Figure 3 displays 4 frames highlighting different aspects of a particle distribution and the corresponding tidal field in an N-body simulation starting from standard CDM initial conditions. The upper lefthand frame shows the particle distribution, at the epoch $a = 0.4$, in a $2.0 h^{-1}$ Mpc thick slice through the centre of the simulation box. The tidal field that corresponds to this mass distribution was determined, via Fourier space calculations, on a $128^3$ grid. A contour plot of the amplitude $|E|$ of this field in the same slice, where we define $|E| \equiv \sqrt{(\sum_{i,j} E_{ij} E_{ij})}$, is shown in the upper righthand frame. The tidal field appears to extend out further and is less compact than the corresponding density field. Also noticable is the considerable strength of the field in the small void around the centre, so instrumental in the subsequent collapse of that small void. Also, we find a strong correlation between the tidal field and the matter distribution, massive clumps of matter corresponding to high peaks in the tidal field. Additional insight into the spatial and orientational structure of the tidal field can be obtained from the plot in both the lower lefthand and righthand frame. Within the same central $x - y$ slice, we determined the principal axes defined by the components $E_{xx}$, $E_{yy}$ and $E_{xy}$ of the tidal field. Tidal compression occurs along the axis corresponding to a positive eigenvalue (left), tidal stretching along the axis representing a negative eigenvalue (right). The solid lines in the lefthand figure are oriented along the direction in which the tidal field causes compression (wrt. an isotropic expansion/contraction), while the ones in the righthand figure are oriented along the direction in which the field causes stretching motions. The size of the edges is proportional to the amplitude of the corresponding (compressional or dilational) components. A conspicuous feature in the compressional map is the almost circular arrangement of edges around massive clumps. Further analysis shows that this arrangement is flattened in a direction perpendicular to the elongation of the mass distribution in the clump. Likewise, we can recognize the most massive clumps in the dilational map, being the sites to which stretching motions are directed. Also notice how strong the compressional component is inside the small void. Moreover, there seems to be a remarkable correlation between structures in the 'tidal compression' map and the large scale matter distribution (compare upper and lower lefthand frames). It is remarkable to see that the correlation is much weaker in the case of the dilational component, in which we can only recognize the most massive clumps. In particular the filaments in the mass distribution appear to represent outstanding features in the compressional map. Most conspicuous is the filament running from the centre to the lower right corner of the slice. This forms a strong indication for an intimate and strong relation between the presence of filaments and the tidal field (also see the recent work of Bond, Kofman & Pogosyan 1995). The presence of filaments seems to be dictated by the presence of a strong tidal field to keep them together !

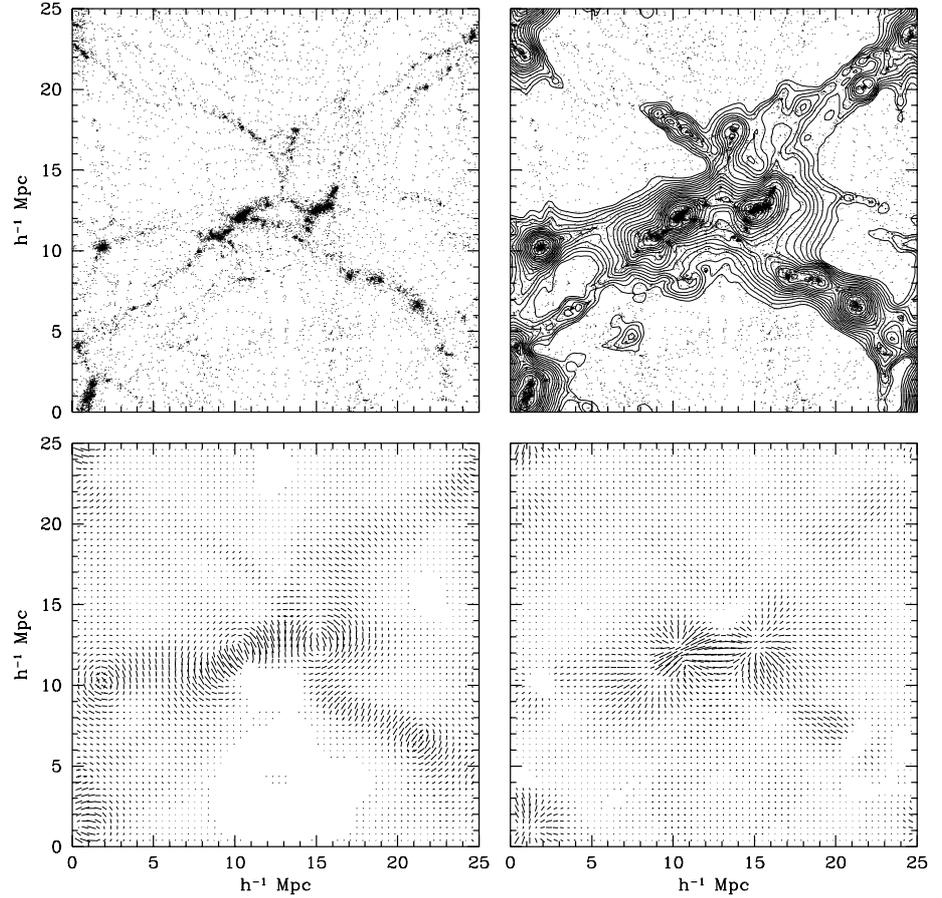

Figure 3.   Tidal field structure in a CDM density field. *Upper left*: Particle distribution in a $2h^{-1}$ Mpc thick slice through the centre of the simulation box. *Upper right*: Strength $|E|$ of tidal field in the same slice, at a resolution of $0.2h^{-1}$ Mpc. Lowest contour at $0.5\langle|E|\rangle$. *Lower left*: Compressional component of tidal field in central $x - y$ plane. Edge directed along compression direction, with a size proportional to the strength of the compression. *Lower right*: Dilational (stretching) component of tidal field in same central plane.

**Acknowledgments.** I gratefully acknowledge my collaborators Arif Babul, Ed Bertschinger and Michiel van Haarlem. This research has been made possible by a fellowship of the Royal Netherlands Academy of Arts and Sciences.